\documentclass{article}
\usepackage[utf8]{inputenc}
\usepackage{authblk}
\usepackage{graphicx}
\usepackage{natbib}
\usepackage{hyperref}

\title{Seven Principles for Effective Scientific Big-Data Systems}
\author[1]{Niall H. Robinson}
\author[2]{Joe Hamman}
\author[3]{Ryan Abernathy}
\affil[1]{Informatics Lab, Met Office, Exeter, UK}
\affil[2]{Climate and Global Dynamics Laboratory, National Center for Atmospheric Research, Boulder, Colorado, US}
\affil[3]{Department of Earth and Environmental Sciences, Columbia University / Lamont Doherty Earth Observatory}
\date{June 2020}

\begin{document}

\maketitle
Thousands of climate scientists have been eagerly awaiting the data from the sixth iteration of the Climate Model Intercomparison Project (CMIP), which is now becoming available online \citep{Balaji_2018}. The data, at a volume that is expected to exceed 20 Petabytes, represent the most detailed predictions ever made about the future of our planet. Buried among the ones and zeros are answers to urgent societal questions, such as which regions should expect more droughts or whether we can expect hurricanes to become more damaging. Yet once the data are released, the waiting will not be over. Scientists will keep waiting...waiting for data to download, waiting for analysis scripts to churn through the files, waiting for that eureka moment when an insightful figure finally appears on their computer screen.

This experience typifies the challenge faced by modern scientific communities. We should be in a golden age of scientific discovery, given that we have more data and more compute power available than ever before. Furthermore, the emergence of machine learning methods which can effectively learn from large datasets holds great promise for scientific research \citep{ReichsteinEtAl2019}. But paradoxically, in many data-driven fields, the eureka moments are becoming increasingly rare. Scientists and their analysis tools are struggling to keep pace with the explosion in the volume and complexity of scientific data. The ``big data revolution'', which has had a major impact across the private industry sector, is failing to similarly empower scientific analysis, which is often more varied, iterative, multidimensional, and interactive than enterprise data science.

For today's scientists, the problem goes beyond the obvious inefficiencies of working with inadequate tools. We argue that the inability to freely explore these datasets creates a pressure to look for ``safe'', expected results---the antithesis of innovative science \citep{liu_2014}. In addition, these stilted workflows drastically interrupt scientists' creative train of thought, limiting discoveries and insights. The insidious growth of this problem is thus a fundamental and widespread blocker to the progress of many modern areas of science.

While compute capacity, memory, and storage have all grown exponentially over the last few decades, network capacity has grown much more slowly. This has shifted our data problem on its axis, away from a problem bound by compute, towards a problem bound by moving data. In enterprise operations, this trend has given rise to the concept of ``data gravity'', where compute is moved to data. In scientific research, which is much more decentralized, the status-quo largely remains a download model, wherein data is moved from central distribution servers to local computing resources.

In addition, the apparent end of Moore's Law scaling means we can no longer simply rely on ever-faster CPUs to facilitate new scientific discoveries \citep{Waldrop_2019}: the main opportunity to continue to accelerate computing performance is via increased parallelism, whether in CPUs, GPUs or more exotic hardware.
Scientific applications which perform highly structured and repetitive operations can easily take advantage of this trend. Running large climate models on high-performance computers is a typical example. However, such simulations only generate even more data for downstream analysis tools to consume. These downstream tools, which must support ad hoc and interactive computations, are traditionally much less effective at leveraging massively parallel computing architectures.

To overcome these challenges, scientists need some sort of a big data ``platform'', which provides both centralized  data storage and a parallel computing framework for analyzing it {\em in situ}.
Numerous such big-data science platforms have been developed, using both open-source and proprietary technologies.
Google Earth Engine \citep{Gorelick_2017} is a prime example of a self-contained platform which excels at the narrow task of geospatial imagery analysis at scale.
Government and scientific organizations have also deployed their own big data platforms with similar ambitions \citep[e.g., for weather and climate data][]{Fiore_2016, Schnase_2017,Raoult_2017}.
These systems employ a wide range of different design principles, from monolithic to highly modular, specialized to general-purpose, and target different underlying computing systems (e.g. traditional on-premises servers, high-performance computers, grid federations, and clouds).

However, there has yet to emerge a consensus approach to designing these systems. We contend that scientific research needs systems which are:
\begin{itemize}
    \item {\em powerful} allowing scientists to apply bespoke analyses to very large amounts of data with a minimal amount of specialist technological expertise,
    \item {\em flexible}, comprising an ecosystem of complementary tools that can be adapted to a wide range of different use cases,
    \item {\em interactive}, recognizing that discoveries are made when scientists can get results quickly, contemplate, and then iteratively refine their calculations,
    \item {\em cost efficient}, only incurring charges whilst computations are being performed, and
    \item {\em sustainable}, facilitating an ongoing maintenance effort that can outlive a single grant-funding cycle.
\end{itemize}
New possibilities for building such systems are now emerging, thanks to a set of rapidly evolving technologies such as cloud computing, container orchestration, automatic parallelism, and thin-client interfaces.

With these challenges in mind and these emerging technologies in hand, here we put forth an opinionated set of principles which can aid the design of scientific infrastructure for big data analytics.
We have implemented and refined these principles over the past few years within the Pangeo Project (\url{http://pangeo.io/}), a grass-roots collaboration between scientists, technologists, and hackers aimed at building community and accelerating research.
The project grew out of the open-source scientific Python community and was inspired by the acute challenges faced by climate scientists, in particular, regarding how to best work with the large multidimensional, gridded datasets produced by satellites and climate models.
The Pangeo approach is, however, being adopted spontaneously in varied fields from astronomy \citep{Barnes_2019} to neuroscience \citep{Rokem_2019} to economics.
Because of this interdisciplinary resonance, we feel these principles merit broadcasting to the greater scientific community.

Pangeo eschews the notion of one monolithic end-to-end platform.
Instead, we aim to cultivate an ecosystem of interoperable tools and architectures which manifest in multiple flavors and instances.
In this article, we spare the technical details of Pangeo, and instead share the principles that we believe are important for effective interactions with scientific data and which serve as a strong foundation future scientific research platforms.

\begin{figure}
    \centering
    \includegraphics[width=0.49\textwidth]{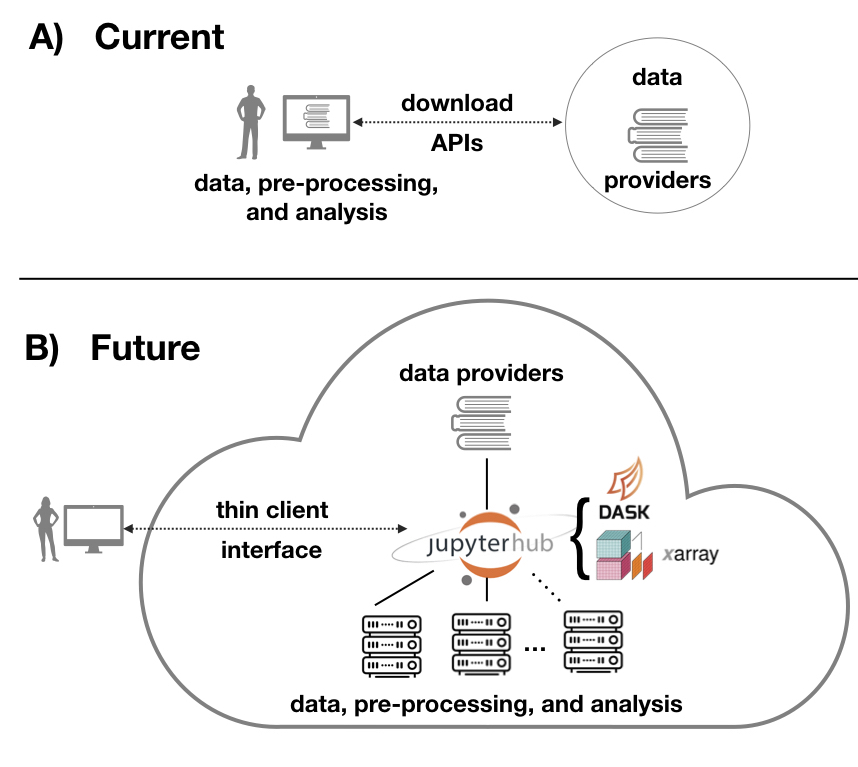}
    \caption{(top) traditional remote data access, where storage is remote and scientist access data through by downloading/copying the archive. (bottom) emerging data proximate paradigm, where analysis, computing and data are collocated.}
    \label{fig:schematic}
\end{figure}

\section*{Separate concerns and specialize late}
It is almost impossible to say what a scientist is going to want to do with their data. Analyses are hand crafted to answer complex questions, and they often need to utilize the latest domain-specific data analysis techniques. Similarly, the technology components that comprise a full system are developing rapidly, with new potentially useful functionality appearing regularly.

As such, any scientific data platform should be readily adaptable, both in terms of the tools available to the end user, and the technologies used in the system. Pangeo achieves this by coupling together a series of unitary components which do one (and only one) thing well - a design principle known as ``separation of concerns.'' For instance, a Pangeo deployment may have a user interface \citep[JupyterLab,][]{Kluyver_2016}, a data model \citep[Xarray,][]{Hoyer_2017}, a parallel job distribution system \citep[Dask,][]{Rocklin_2017}, a system for managing resources \citep[Kubernetes,][]{Rensin_2015}, a raw data storage system (AWS S3), and a broker to provide analysis ready versions of the raw data (Intake). This modularity makes it practical to adapt the individual components, as long as they still fulfill the same purpose.

For the system to be adaptable to different use cases, it is important that as few components as possible are affected by domain specific design decisions - a principle which we term ``specializing late''. For instance, if the data store component needlessly assumes that data will be stored in a geospatial data forms, it prohibits the system from being used later for astrophysics.

These architectural principles of {\em separation of concerns} and {\em specializing late} seem to be regularly disregarded in the creation of scientific platforms. Sometimes, this is simply because of the rush to implement a system that works. Often it is in the pursuit of optimization: the temptation is to finely tune a system to do all the things that an end consumer could possibly want to do...only to have a user want it to do something else. If a system has been over-engineered to optimize the performance of a very specific use case, it can be much more difficult to extend. These kind of finely tuned but brittle systems can prove useful for non-research data analysis, where use cases are relatively predictable and performance is of the utmost importance. However, we maintain that scientific analysis is inherently variable and necessitates a flexible system.

Keeping components of the systems as generic as possible future proofs them, allowing the thin layer of domain specific functionality built on top of them to be adapted to new use cases. For instance, in Pangeo, analysts interact with objects which represent earth science data (Xarray or Iris), and these systems invoke more generic operations in other components, such as NumPy \citep{van_2011} and Dask \citep{Rocklin_2017}. This is why Pangeo, initially deployed for climate science, is easily repurposed for other fields, such as genomic and astronomical data analysis.

Finally, specializing late encourages data system developers to contribute to the ``upstream'' tools on which their platform depends.
Funding for Pangeo has nearly all been specifically to develop tools of use to geoscientists.
However, in the process, he project has been able to contribute significantly to more general-purpose open-source tools such as NumPy, Dask, and Jupyter, which serve a much broader user base.
Ironically, such general-purpose tools have struggled to attract federal funding support commensurate with their impact \citep{Eghbal_2018}, a probable consequence of the disciplinary divisions at funding agencies.
The {\em specialize late} principle allows disciplinary cyberinfrastructure providers, whether in geoscience or genomics, to justify funneling much needed developer effort to these essential projects.

\section*{Co-locate compute and data}

As scientific datasets grow towards the petabyte scale, they become harder and harder to move over the internet.
Instead we need to move computations closer to the stored data, to a remote computer coupled with high bandwidth to the data store.
Interacting with remote computing systems has traditionally meant using a UNIX-style command line application, a barrier in terms of productivity and accessibility for interactive data analysis.
However, the emergence of new paradigms for remote interaction, focused on the web browser rather than the terminal, can remove these barriers.
Pangeo makes use of JupyterLab \citep{Kluyver_2016}, a web-based interactive development environment for Jupyter Notebooks, code, and data,
which provides a common interface to both local and remote computing systems. 
Scientists interact with JupyterLab in their web browser, which sends commands and queries via the internet to compute resources co-located with data. Scientists who do computing this way only require a ``thin-client''---a minimally powerful workstation with a web browser---on their desktop.
Indeed, Pangeo can also be used on a tablet or smartphone.

This sort of data-centric configuration can be realized on national-level HPC systems, which typically provide multi-petabyte high-performance filesystems accessible from compute nodes. Such systems are commonly used for generating and storing scientific data. While HPC systems typically prioritize batch-style computations, there is no fundamental reason why they can't be adopted for interactive workloads. These kind of highly optimized, tightly coupled systems offer great performance. However, they are not accessible to all scientific data consumers, and they are notoriously highly secure and locked down.

The commercial cloud, with its nearly infinite compute and storage capacity, is another viable choice for bringing compute to data.
In contrast with HPC, cloud users are free to customize their software environment, security protocols, and computing hardware to their liking.  However, there remain challenges to wider adoption of cloud computing in science, primarily the shift from capital expenditure funding to ongoing operational expenditure. Beyond this challenge of administering funding, moving to operational expenditure offers a great advantage, as the user only pays for what they use.

\section*{Compute in parallel and scale elastically}
Parallel computing is common in traditional HPC simulation, where technologies like MPI are employed to distribute large calculations over many compute nodes, decreasing the time to solution for large computational domains.
And when dealing with multi-terabyte or petabyte scale datasets, parallelization of some sort is the only way to achieve the speed necessary for interactive analysis.

However, parallel computing is not yet widespread in scientific data analysis, even though many data analysis problems are trivial to parallelize.
This is largely a software limitation---legacy analysis tools were simply not designed for distributed parallelism.
Widely used enterprise ``big data'' tools such as Hadoop \citep{White_2012} or Spark \citep{Zaharia_2016} do leverage parallelism,
offering a suite of opinionated data processing operations suitable for oft-repeated analyses based on the map-reduce paradigm.
However, these systems have only been of limited use for scientific analysis \citep{Schnase_2017}, largely due to the nature of scientific data processing, which is typically more highly dimensional, more ad hoc, and more complex. 
On the other hand, specialized scientific big-data tools like Google Earth Engine may excel at one use case (geospatial imagery processing) while being impossible to adapt to another (bioinformatics).
Put another way, science does not need a train, it needs an all terrain vehicle.

An appealing alternative has emerged in tools such as Dask \citep{Rocklin_2017}, Modin \citep{petersohn_2020}, and Vaex \citep{Breddels_2018}, which mimic the familiar interface of general-purpose scientific data analysis libraries like NumPy and Pandas, while, under the hood, executing calculations in parallel.
In Pangeo, Dask is used to distribute calculations across many nodes of an HPC or cloud-based cluster, or across the available cores on a single computer.

In contrast to traditional simulation, data analysis workloads are highly variable in time.
Often researchers process some data, produce a figure, and spend a few minutes staring at it before performing another calculation.
In this scenario, it's wasteful to pre-allocate a fixed number of compute nodes, which will remain idle for much of the time.
Instead, we advocate taking advantage of ``elastic'' scaling, in which compute nodes are rapidly allocated and deallocated (on the order of seconds or minutes) in response to user activity.
Elastic scaling is only possible in very large compute facilities, which can absorb individual user volatility while still making efficient use of their resources.
Large HPC centers traditionally choose to optimize for maximum system utilization by queuing jobs on batch schedulers.
Conversely, commercial cloud platforms optimize for availability and are an ideal environment for elastic scaling, given their vast size and diverse user base.
Most cloud providers charge for computing by the minute, meaning that it costs the same to use one computer for 1000 minutes as it does to use 1000 computers for one minute.
This sort of on-demand elastic scaling has transformative possibilities for scientific data analysis, allowing scientists to analyses orders of magnitude faster for little extra cost.

We have deployed Pangeo in elastic scaling mode in both HPC and cloud environments.
When a user executes an analysis, a Pangeo compute cluster can automatically scale to hundreds of compute nodes and distribute the work before relinquishing the compute nodes. Thanks to Dask, user does not have to explicitly parallelize their work---they can execute the same high-level scientific analysis they are used to, but they get their answer faster. Working this way facilitates a more interactive relationship with data, leading to more creativity and new discoveries.

\section*{Analyze data lazily}
We can also make sure that our system gets answers faster (and cheaper) by doing the minimal amount of analysis possible - so called ``lazy evaluation''. Traditionally, expensive calculations are often performed once and stored on disk for later inspection and analysis. However, this approach comes with multiple downsides. It is often cost inefficient, as storage can be more expensive than re-computation, especially in the case of sparsely accessed data. It also obscures the provenance of the data. Finally, it is a brittle process, as it is hard to change the nature of the analysis.
Instead, we promote maintaining canonical, base-level datasets along-side ``lazy'' derived datasets, which access and process the canonical dataset on-demand. This was previously impractical as processing times were too long for on-demand data products; however, it becomes a feasible proposition with elastic scaling and parallel computing.

For instance, calculating the difference between historical surface temperature and satellite observations is an inexpensive calculation which results in a large data set. Pangeo allows users to subtract these two fields to create a ``lazy'' data object. This operation produces an object that represents the resultant field, but which encapsulates the latent calculation, executing it only when the data itself is accessed. For instance, if the user then chooses to look at the values over London, only the necessary data is pulled from the data store, and only that calculation is performed.

\section*{Publish analysis-ready data}
Just the act of accessing a dataset is now often a great hurdle to doing science. Big scientific datasets are commonly stored as many separate bite-sized chunks so that they can be accessed in parallel and are fault tolerant (for example, one data file per day of a global satellite product). Reconstructing meaningful data objects from these chunks is an increasingly non-trivial task: metadata must be defined for each chunk and then data concatenated to create the true, meaningful data object, so-called ``analysis-ready'' data.
Tools such as Xarray or Iris attempt to solve this problem by collecting metadata from multiple chunks before automatically representing them as a composite data object. However, this approach does not currently scale well, and the logic for constructing composites often requires expert knowledge.

In reality, the step of finding data and cajoling it into a representative meta-object to work with is too often a huge burden on the data consumers. Moreover, this difficult task is repeated by every consumer who uses a dataset. The lack of analysis-ready data forces scientists into inefficient patterns, such as manually iterating over chunks.

Within Pangeo, we have employed multiple approaches to lower the cognitive burden for accessing large datasets.
One is to explore new data containers such as Parquet and Zarr, which are optimized for the storage of very large datasets across many individual files.
The I/O libraries for these formats can then shoulder the burden of assembling all the individual pieces into a single, coherent data object, using the lazy-access principle described above.
A complimentary approach is to use a data-broker layer, such as Intake, which mediates between the data files and the composite dataset object a scientist uses in their analysis session. The complex task of defining how to load and combine chunks is done once by the person who knows the most about the data: the data generator. This recipe is captured in a catalog ``driver'' which is published to allow access to the dataset. The scientist can then simply install the driver for a particular dataset before loading a manicured, ready to use representation with one line of code. They do not need to pause to think about file paths or formats, let alone combining chunks.

\section*{Build on open infrastructure}
Big data systems, whether hosted in the public cloud or elsewhere, require significant infrastructure to be effective. This comprises the hardware and software to: store large amounts of data; efficiently and dynamically scale computing resources; and connect computing to data via high-bandwidth networking. While we have found that commercial cloud computing systems provide many of the necessary building blocks to assemble effective scientific big data platforms, care must be clearly be taken to avoid vendor lock-in, that is, becoming dependent on a single big tech company like Google, Amazon or Microsoft. This can be ameliorated by adopting open standards and building on common interfaces.  For instance, Kubernetes is an open source cloud-native system which mediates between running processes and infrastructure, and it is supported by all the major cloud providers. By taking advantage of this system, the Pangeo project has run data platforms on AWS, Microsoft Azure, Google Compute Cloud and Alibaba.



\section*{Build federations, not monoliths}
The flexibility of the systems architecture proposed here means we can divide up our data platforms and locate different sections in different places. Pangeo deployments can be created by implementing an automated recipe which defines the entire interconnected system---a technique known as Infrastructure as Code (IaC). This is a powerful advantage for several reasons.

Firstly, whilst cloud computing clearly offers some profound new functionality, it is unreasonable and unaffordable to propose that all scientific data can be moved to the cloud, at least in the near future. However, our recipe approach means we can now move a portion of the data platform to the data archive. For instance, a user based in the cloud could analyze one data object, which transparently invokes operations on the archive compute cluster, before seamlessly analyzing another data object which is hosted in the cloud. From the user’s perspective the only difference would be that the latter might process quicker, by taking advantage of scalable cloud compute.

Secondly, this flexibility allows us to gracefully apportion costs to different stakeholders. Data generators increasingly want to make their data available and useful to outside organizations and commercial companies so they can translate data into social and economic impact. However, this leads to a tension: How can the data providers empower third parties to build customized data-driven applications, without having to pay for the third party's computing costs? This can be particularly problematic for publicly funded data generators, who want their data to benefit the general economy, but without unintentionally subsidizing private industry. Our recipe approach allows these third parties to mirror portions of the platform on their own cloud computing account. They thereby shoulder any cost which they should subsequently recoup from monetizing the derived data products.

In general, we envisage a move away from the false dichotomy of “on premise data platforms” and “cloud computing data platforms”, towards a federated ecosystem of data platforms, which are interconnected based on these common interfaces.

\section*{Conclusion}
In talking to colleagues from diverse research fields, from climate science to astronomy to neuroscience, we see a need for a seismic shift in the way the scientific community interacts with large datasets.
The principles described here can serve to guide the construction of robust data analysis environments which can meet the challenges of modern data-driven scientific research.
We have attempted to implement these principles in the Pangeo Project, which we believe has created a solid foundation for all manner of workflows.
With such a foundation in place, we can now focus on higher-level concepts, such as building effective user interfaces and visualizations and exploring new algorithms to extract more information from these datasets.
The end goal of these efforts is to move scientists away from the drudgery of processing their big data, and back to intuitive and productive workflows, thereby accelerating scientific progress and providing answers to our important societal questions.

\bibliographystyle{unsrt}  
\bibliography{references}

\begin{thebibliography}{10}

\bibitem{Balaji_2018}
V.~Balaji, K.~E. Taylor, M.~Juckes, B.~N. Lawrence, P.~J. Durack,
  M.~Lautenschlager, C.~Blanton, L.~Cinquini, S.~Denvil, M.~Elkington,
  F.~Guglielmo, E.~Guilyardi, D.~Hassell, S.~Kharin, S.~Kindermann, S.~Nikonov,
  A.~Radhakrishnan, M.~Stockhause, T.~Weigel, and D.~Williams.
\newblock Requirements for a global data infrastructure in support of cmip6.
\newblock {\em Geoscientific Model Development}, 11(9):3659--3680, 2018.

\bibitem{ReichsteinEtAl2019}
Markus Reichstein, Gustau Camps-Valls, Bjorn Stevens, Martin Jung, Joachim
  Denzler, Nuno Carvalhais, and Prabhat.
\newblock Deep learning and process understanding for data-driven earth system
  science.
\newblock {\em Nature}, 566(7743):195–204, Feb 2019.

\bibitem{liu_2014}
Zhicheng Liu and Jeffrey Heer.
\newblock The effects of interactive latency on exploratory visual analysis.
\newblock {\em IEEE transactions on visualization and computer graphics},
  20(12):2122--2131, 2014.

\bibitem{Waldrop_2019}
{Waldrop, M. Mitchell}.
\newblock The chips are down for moore’s law.
\newblock {\em Nature}, 530:144--147, 2016.

\bibitem{Gorelick_2017}
Noel Gorelick, Matt Hancher, Mike Dixon, Simon Ilyushchenko, David Thau, and
  Rebecca Moore.
\newblock Google earth engine: Planetary-scale geospatial analysis for
  everyone.
\newblock {\em Remote Sensing of Environment}, 202:18 -- 27, 2017.
\newblock Big Remotely Sensed Data: tools, applications and experiences.

\bibitem{Fiore_2016}
S.~{Fiore}, M.~{Płóciennik}, C.~{Doutriaux}, C.~{Palazzo}, J.~{Boutte},
  T.~{Żok}, D.~{Elia}, M.~{Owsiak}, A.~{D'Anca}, Z.~{Shaheen}, R.~{Bruno},
  M.~{Fargetta}, M.~{Caballer}, G.~{Moltó}, I.~{Blanquer}, R.~{Barbera},
  M.~{David}, G.~{Donvito}, D.~N. {Williams}, V.~{Anantharaj}, D.~{Salomoni},
  and G.~{Aloisio}.
\newblock Distributed and cloud-based multi-model analytics experiments on
  large volumes of climate change data in the earth system grid federation
  eco-system.
\newblock In {\em 2016 IEEE International Conference on Big Data (Big Data)},
  pages 2911--2918, Dec 2016.

\bibitem{Schnase_2017}
John~L. Schnase, Daniel~Q. Duffy, Glenn~S. Tamkin, Denis Nadeau, John~H.
  Thompson, Cristina~M. Grieg, Mark~A. McInerney, and William~P. Webster.
\newblock Merra analytic services: Meeting the big data challenges of climate
  science through cloud-enabled climate analytics-as-a-service.
\newblock {\em Computers, Environment and Urban Systems}, 61:198 -- 211, 2017.
\newblock Geospatial Cloud Computing and Big Data.

\bibitem{Raoult_2017}
Baudouin Raoult, Cedric Bergeron, Angel~L{\'o}pez Al{\'o}s, Jean-No{\"e}l
  Th{\'e}paut, and Dick Dee.
\newblock Climate service develops user-friendly data store.
\newblock {\em ECMWF}, pages 22--27, 2017.

\bibitem{Barnes_2019}
Will Barnes, Chun Ming~Mark Cheung, and Monica Bobra.
\newblock The sun at scale: Interactive analysis of high resolution euv imaging
  data on hpc platforms with dask.
\newblock In {\em AGU Fall Meeting 2019}. AGU, 2019.

\bibitem{Rokem_2019}
Ariel Rokem.
\newblock Rokem research, Oct 2019.

\bibitem{Kluyver_2016}
Thomas Kluyver, Benjamin Ragan-Kelley, Fernando P{\'e}rez, Brian~E Granger,
  Matthias Bussonnier, Jonathan Frederic, Kyle Kelley, Jessica~B Hamrick, Jason
  Grout, Sylvain Corlay, et~al.
\newblock Jupyter notebooks-a publishing format for reproducible computational
  workflows.
\newblock In {\em ELPUB}, pages 87--90, 2016.

\bibitem{Hoyer_2017}
Stephan Hoyer and Joe Hamman.
\newblock xarray: Nd labeled arrays and datasets in python.
\newblock {\em Journal of Open Research Software}, 5(1), 2017.

\bibitem{Rocklin_2017}
Matthew Rocklin.
\newblock Dask: Parallel computation with blocked algorithms and task
  scheduling.
\newblock In Kathryn Huff and James Bergstra, editors, {\em Proceedings of the
  14th Python in Science Conference}, pages 130 -- 136, 2015.

\bibitem{Rensin_2015}
David~K Rensin.
\newblock {\em Kubernetes-scheduling the future at cloud scale}.
\newblock O'Reilly, 2015.

\bibitem{van_2011}
Stefan Van Der~Walt, S~Chris Colbert, and Gael Varoquaux.
\newblock The numpy array: a structure for efficient numerical computation.
\newblock {\em Computing in Science \& Engineering}, 13(2):22, 2011.

\bibitem{Eghbal_2018}
Nadia Eghbal.
\newblock Roads and bridges: the unseen labor behind our digital
  infrastructure. ford foundation, 2018.

\bibitem{White_2012}
Tom White.
\newblock {\em Hadoop: The definitive guide}.
\newblock " O'Reilly Media, Inc.", 2012.

\bibitem{Zaharia_2016}
Matei Zaharia, Reynold~S. Xin, Patrick Wendell, Tathagata Das, Michael
  Armbrust, Ankur Dave, Xiangrui Meng, Josh Rosen, Shivaram Venkataraman,
  Michael~J. Franklin, and et~al.
\newblock Apache spark: A unified engine for big data processing.
\newblock {\em Commun. ACM}, 59(11):56–65, October 2016.

\bibitem{petersohn_2020}
Devin Petersohn, William Ma, Doris Lee, Stephen Macke, Doris Xin, Xiangxi Mo,
  Joseph~E. Gonzalez, Joseph~M. Hellerstein, Anthony~D. Joseph, and Aditya
  Parameswaran.
\newblock Towards scalable dataframe systems, 2020.

\bibitem{Breddels_2018}
{Breddels, Maarten A.} and {Veljanoski, Jovan}.
\newblock Vaex: big data exploration in the era of gaia.
\newblock {\em A \& A}, 618:A13, 2018.

\end{thebibliography}

\end{document}